 \documentclass[12pt,draftcls,onecolumn]{IEEEtran}

\pdfoutput=1 

\usepackage{cite} \usepackage{graphicx} \usepackage{subfigure} \usepackage{url} \usepackage{amsmath,amssymb} \usepackage{bm} \usepackage{hyperref}
\usepackage{color}
\usepackage{amssymb}  %
\usepackage{bm}
\usepackage{amsmath}
\usepackage{algorithmic}
\usepackage{algorithm}

\hypersetup{
	pdfauthor={Fabio Ricciato, Paolo Castiglione},
	pdftitle={xxxx}
}

\newcommand{\eqdef}{ \stackrel{\textrm{\tiny def}}{=}}

\begin{document}

\title{Pseudo-random Aloha for Enhanced Collision-recovery in RFID\footnote{This work has been submitted to the IEEE for publication. Copyright may be transferred without notice, after which this
version may no longer be accessible.}}

\author{Fabio Ricciato\authorrefmark{1}, Paolo Castiglione\authorrefmark{2} \authorblockA{\\\authorrefmark{1}Universit\`{a} del Salento, Lecce, Italy\\ \authorrefmark{2}Telecommunications Research Center Vienna (FTW), Vienna, Austria}}

\maketitle              

\begin{abstract}
In this letter we motivate the need to revisit the MAC protocol used in Gen2 RFID system in order to leverage receiver structures with Collision Recovery capabilities at the PHY layer. To this end we propose to consider a simple variant of the Framed Slotted Aloha with pseudo-random (deterministic) slot selection as opposite to the classical random selection. Pseudo-random access allows naturally to implement Inter-frame Successive Interference Cancellation (ISIC) without changing the PHY modulation and coding format of legacy RFID standard.
By means of simulations we show that ISIC can bring 20-25\% gain in throughput with respect to traditional intra-frame SIC.
Besides that, we elaborate on the potential of leveraging pseudo-random access protocols in combination with advanced PHY techniques in the context of RFID applications.
\end{abstract}

\section{Introduction and motivations}\label{sec:intro}
{Passive RFID systems are being applied in a growing and diverse range of real-world applications}.
One important class of RFID applications addresses   \emph{delay-sensitive dense scenarios}, where a large number of RFID Tags must be read within the minimum possible time by a Reader device --- think of, e.g., a shopping cart carrying hundreds of tagged items traversing a RFID reading gate.
In such applications the primary goal is to minimize the total reading time, i.e., the time needed to read all Tags, or equivalently to maximize the \emph{reading throughput}.
The attractiveness for passive RFID technology ultimately resides in the ultra-low cost nature of the Tags, therefore any new proposal oriented to increase reading throughput should not require increased complexity or additional resources (e.g. processing power) on the Tag side.
In this work we claim that a considerable throughput gain can be achieved with only a minor change --- actually a \emph{simplification} --- of the standard MAC scheme.
In the proposed solution the MAC protocol overhead on the Tag side is \emph{reduced} while reading performance are increased.  The cost to be paid is somewhat increased complexity of the reader receiver structure, which is not a cost-critical element of the whole RFID system.

The most popular RFID standard, namely EPC Global Generation-2 specifications (``Gen2" for short) \cite{gen2}, foresees an interrogation scheme based on Framed Slotted Aloha.
The design of the MAC scheme in the Gen2 protocol is based on the assumption that tag collisions \footnote{We refer hereafter exclusively to ``tag collisions" (uplink collisions), i.e.  when two or more tags collide on the same slot at the reader receiver. Reader collisions (downlink collisions) caused by the contemporary transmission of multiple readers are not relevant to this work.}
 are always destructive events, causing the loss of all colliding messages. For this reason, a sort of Collision Avoidance (CA) mechanism is foreseen by Gen2, based on the preliminary exchange of shorter 16-bit messages in each slot for channel reservation: the \texttt{RN16} in uplink and \texttt{ACK} in downlink. Therefore, collisions can occur only between (two or more) short RN16 messages, not on the 128-bit message of interest.

The assumption that collisions are always destructive is a gross limitation of Gen2.
First, even the simplest receiver will decode at least the strongest of the colliding messages provided that the power gap  with the other(s) weaker signal(s) is sufficiently large: this is commonly known as ``capture effect" and was  observed experimentally in \cite{Buettner11}.
Once that the strongest signal has been decoded, the receiver can cancel it from the total received signal and then attempt to decode the second strongest signal, and so on. This method, known as Successive Interference Cancellation (SIC), allows to resolve multiple colliding signals at the PHY layer under certain conditions.
Generally speaking, more advanced PHY receiver structures can be adopted to recover multiple colliding messages, exploiting more or less sophisticated ``Collision Recovery" (CR) techniques --- also referred to as ``multi-user detection"--- such as those proposed recently in \cite{rupp-collisionrec} and \cite{scaglione} specifically for backscattered RFID signals. However none of such works has dwelled into the implications \emph{at the MAC layer} of doing so. This is the focus of this letter.

When CR is implemented at the PHY layer, \emph{collisions cease to be a problem and become an advantage} as they increase the reading throughput.
It is evident the fundamental conflict between the  Collision \emph{Recovery} techniques available at PHY layer, designed to \emph{exploit} collisions, and the  Collision \emph{Avoidance} mechanism mandated by the Gen2 standard in the MAC, based on \texttt{RN16}/\texttt{ACK} exchange, designed to \emph{prevent} them.
Simply put, it is necessary to change the Gen2 MAC protocol in order to leverage the potential gain of any CR technique. 
In the rest of this letter we present for the first time the principles of a novel MAC scheme that is extremely simple --- both conceptually and in terms of implementation, anyway simpler than the current Gen2 --- and at the same time ``PHY friendly". Our proposal is basically a variation of the traditional Framed Slotted Aloha, but with a very fundamental difference: \emph{randomization is replaced by pseudo-randomization}.
In the proposed scheme, slot selection within each frame is not random, as in the classical scheme, but instead based on a \emph{deterministic} (pseudo-random) function of the message to be transmitted and of the frame number.
This (apparently small) detail  enables a more effective cooperation between the MAC and PHY layers which translates into througput gain.
More specifically, it allows to cancel the decoded messages from the received signals in the previous frames, i.e. to implement Inter-frame SIC (ISIC).

We remark that our pseudo-random access scheme  does not make any assumption about the underlying PHY implementation: it \emph{allows} to exploit CR techniques if present,  but does \emph{not require} them to be in place. Also, differently form other proposals \cite{cassini07} our scheme does not require any change to the PHY layer and can be applied directly to the legacy modulation and encoding format of Gen2.




\section{Protocol Description}\label{sec:scenario}
\subsection{Overview}
%

The goal of this letter is to indicate a research direction and draw attention towards the potential of pseudo-random (as opposite to random) access protocols in conjunction with advanced PHY techniques,  rather than working out a specific implementation.  For this reason we describe here the basic version of the proposed scheme, focusing on the central features and core concepts\footnote{
More sophisticated variants can be derived from the basic version by considering more involved strategies for scheduling (re)transmission attempts, e.g. with dynamic adaptation of frame size or frameless binary splitting methods such as those elaborated in \cite{binarytree}.}.

\begin{figure}[tb]
\centering
\includegraphics[width= 0.8\textwidth]{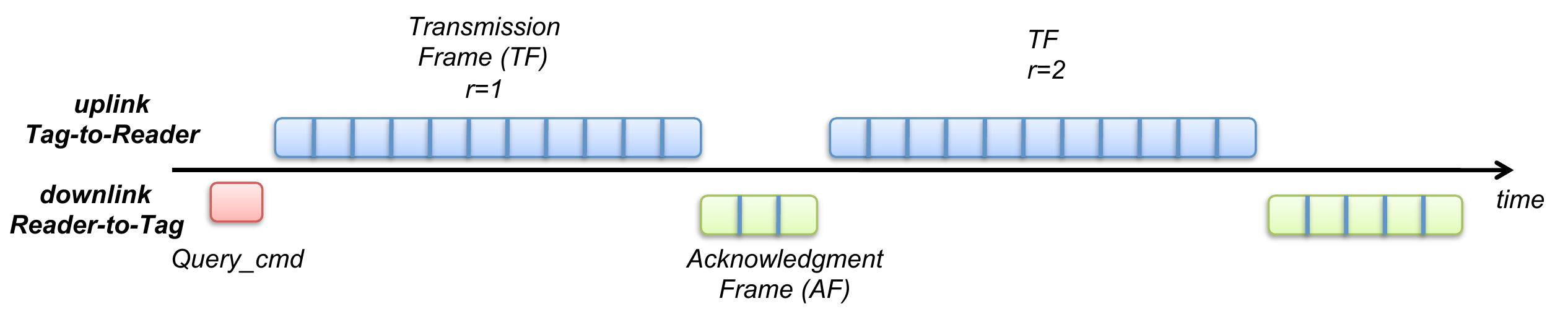}

\caption{Protocol timeline with fixed-size Transmission Frames
and variable-size Acknowledgment Frames.  }
\label{fig:proto}
\end{figure}

From the discussion in the previous section it is evident that the preliminary channel reservation phase, i.e. the \texttt{RN16} and \texttt{ACK} messages  ``playing against" CR must be dropped. In this way the residual protocol reduces to (a variation of) the standard Frame Slotted Aloha.
For the sake of simplicity  we consider here only fixed-size frames with a constant number of slots (recall note$^2$).
We also assume that the Reader acknowledges the messages that have been correctly received (i.e., with successful CRC) at the end of each frame,
in order to inhibit  the acknowledged Tags from retransmitting again in the following frames of the same Query Cycle.
The resulting protocol timeline is represented in Fig.\,\ref{fig:proto}.

The initial \texttt{Query Command} (\texttt{QC}), sent by the Reader, starts a new Reading Cycle: it resets all Tag flags,  and broadcasts the relevant initializazion parameters (e.g. coding parameter, frame size).
The \texttt{QC} message is followed by a sequence of alternating Transmission Frames (TF) in uplink and Acknowledgment Frames (AF) in downlink.
Each TF  is divided into $K$ ``transmission slots" of fixed duration sufficient to accommodate the 128-bit Tag ID.
The AF is of variable length: it consists of a fixed-format preamble plus a variable number of ``acknowledgment slots". 
Each slot is meant to acknowledge the succesful reception of one specific Tag ID in the previous TF: in the simplest option it contains a mere repetition of the decoded 128-bit ID.

At each frame, the generic Tag selects one slot according to the pseudo-random function described later, and transmits its 128-bit ID therein.
It then listens  to the following AF \footnote{It is convenient to sort the acknowledged ID in each AF so as to allow a generic tag to suspend listening and switch to sleep mode as soon as the acknowledged ID exceeds its own ID.}:
if  acknowledged it will set its flag and leave the Reading Cycle, otherwise it will retry at the next frame.

\newcommand{\myxi}{\mathbf{\boldsymbol{x}}_{i}}

\subsection{Pseudo-random slot selection}\label{sec:selalg}

In order to described the principle of pseudo-random slot selection it is convenient to introduce the following notation:
\begin{itemize}
\item $I$ is the variable denoting the total number of active Tags and $i \in \{1,2 \ldots I \}$ is the Tag index.
\item $\myxi$  is the binary ID of tag $i$ composed of $B=128$ bits.
\item $R$ is the variable denoting the total number of TF in a given Reading Cycle, until all Tags are succesfully decoded, and $r \in \{1,2 \ldots R \}$ is the TF index.
\item $K$ is the number of slots in a generic TF and  $k \in \{1,2 \ldots K \}$ is the slot index within a generic TF.
\item $s_{i,r}\in\left\{ 1,\ldots,K\right\} $ is the slot index selected by the $i$th Tag in the $r$th TF.
 \end{itemize}

We consider a pseudo-random function $h(\myxi,r): (\myxi,r) \mapsto s_{i,r}$
 that takes in input the 128-bit Tag ID and the frame index $r$ and returns a value in $[1,R]$ 
In the generic $r$th TF,  Tag $i$ uses this function to select the slot to transmit, i.e. $s_{i,r} = h(\myxi,r)$.

In principle any pseudo-random function can serve the purpose, provided that it can be implemented in ultra-low cost Tag with limited processing and memory resources.
In the following simulations we have used a simple hashing method: the binary sequence associated to the frame index $r$ is concatenated to $\myxi$, and the resulting string is permuted using a pseudo-random interleaver. The latter is then fed to a standard CRC-16 and the first  $\log_2{K}$ bits are picked to indicate the slot position $s_{i,r}$. {In practice the pseudo-random slot positions can be pre-computed and stored in the Tag memory to save processing resources on the Tag side.}

\subsection{Inter-frame Successive Interference Cancellation}\label{sec:isic}

The pseudo-random function $h(\cdot)$ is known at the Reader.
Upon successful detection of a Tag ID $\myxi$, the Reader can identify the slot positions where the same ID was transmitted in all past frames, and therefore can cancel it from the signals received therein. Cancellation can trigger the decoding of other ID that in turn can be cancelled from all past frames, and so on.
In other words, an important  advantage of deterministic (pseudo-random) slot selection is that SIC can be implemented  across different frames and \emph{without any PHY layer amendment}. This is in contrast to \cite{cassini07} and follow-up works, where additional bits are inserted in the preamble in order to carry the information about the slot position.

\begin{figure}[tb]
\centering
\includegraphics[width= 0.8\textwidth]{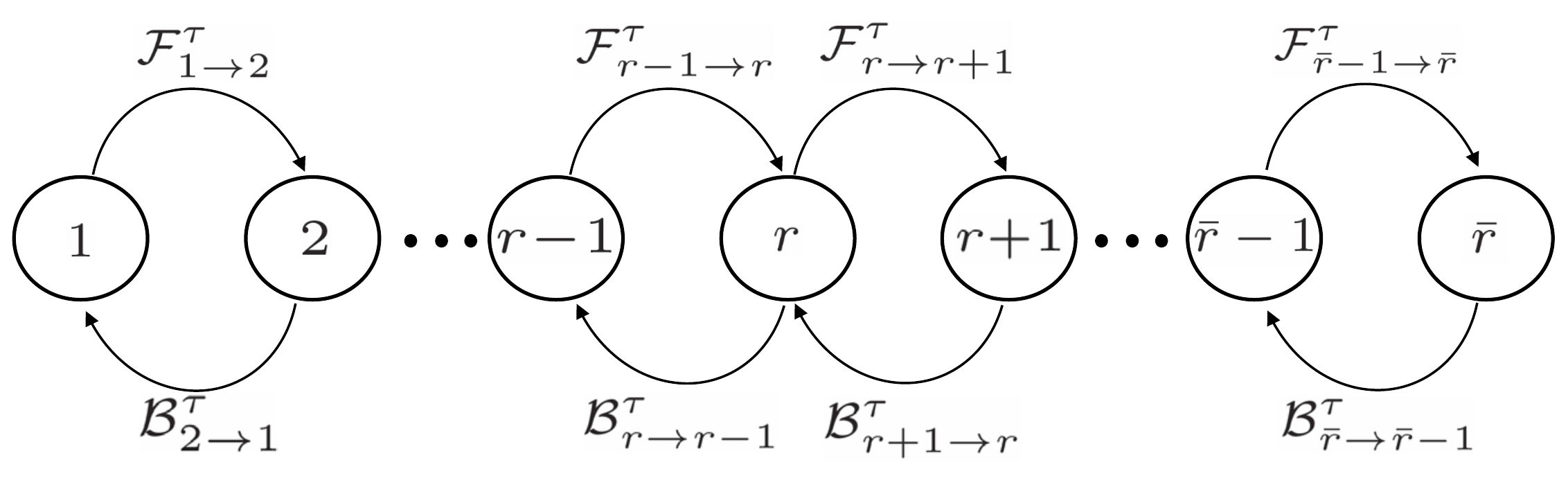}

\caption{ISIC message-passing graph. }
\label{fig:ISIC}
\end{figure}

 Here, we propose an algorithm to perform an \emph{exhaustive} ISIC. Our algorithm  can be represented on the \emph{message-passing graph} shown in Fig.\,\ref{fig:ISIC}, where the residual signals after SIC at each frame $r$ are associated to each node $r$. The symbols $\mathcal{B}_{r+1\rightarrow r}^{\tau}$ and $\mathcal{F}_{r-1\rightarrow r}^{\tau}$ represent respectively the backward messages from node $r+1$ to node $r$ and the forward messages from node $r-1$ to node $r$ at iteration $\tau$. These messages contain tag signals (detected in other frames) that have not been canceled yet from the residual signals of frame $r$. Once the message coming either from the neighboring node $r-1$ or from $r+1$ is received at node $r$, the contained signals can be canceled from the residual signals of frame $r$. Thus, the detection of more tags on the new residual signals of frame $r$ becomes possible and the intra-frame SIC can be repeated. The above described ISIC operation at node $r$ is denoted by the function
 $\mathcal{D}_{r}^{\tau}=\mathrm{ISIC}(\mathcal{B}_{r+1\rightarrow r}^{\tau} \, \mathrm{or} \, \mathcal{F}_{r-1\rightarrow r}^{\tau})$,
where the set $\mathcal{D}_{r}^{\tau}$ contains the new detected signals of frame $r$. Each iteration of the algorithm $\tau$ entails a sequential generation of backwards messages  $\mathcal{B}_{r\rightarrow r-1}^{\tau}$, that are passed from node $\bar{r}$ to node $1$, followed by a sequential generation of forwards messages $\mathcal{F}_{r\rightarrow r+1}^{\tau}$, that are passed from node $1$ to node $\bar{r}$. The detailed algorithm, along with the rigorous definition of the messages, is illustrated in Algorithm \ref{alg:ISIC}.

\begin{algorithm}[!t]
\caption{Exhaustive ISIC message-passing algorithm}
\label{alg:ISIC}
\scriptsize

\begin{algorithmic}[0]
%
%
\STATE{\label{ln:Init}\underline{Initialization:} SIC detection at current frame $r=\bar{r}$;\\the set $\mathcal{D}_{\bar{r}}^{\tau}$ contains the detected tag signals;\\$\tau=1$ is the ISIC iteration index;}
%
%
\IF{$\mathcal{D}_{\bar{r}}^{\tau} \neq \varnothing$}
\STATE{initialize detected tags counter: $D(\tau)=|\mathcal{D}_{\bar{r}}^{\tau}|$};
\STATE{initialize backward message (set): $\mathcal{B}_{\bar{r}\rightarrow\bar{r}-1}^{\tau}\equiv\mathcal{D}_{\bar{r}}^{\tau}$;\\other messages $\mathcal{B}_{r\rightarrow r-1}^{\tau}$ and $\mathcal{F}_{r\rightarrow r+1}^{\tau}$ initialized as empty sets $\varnothing$;}
\ELSE  \STATE{\bf{exit}}
\ENDIF
\STATE{\label{ln:ALG}\underline{Start ISIC algorithm:}}
\WHILE{$D(\tau)>0$}

\STATE{\label{ln:back}initialize: $D(\tau+1)=0$;\\ \underline{Backward message-passing:}}
\FOR {$r = \bar{r}-1\to 2$}
\STATE{\label{ln:backwards}detection: $\mathcal{D}_{r}^{\tau}=\mathrm{ISIC}(\mathcal{B}_{r+1\rightarrow r}^{\tau})$;\\backward message:\\$\mathcal{B}_{r\rightarrow r-1}^{\tau}=\mathcal{D}_{r}^{\tau}\cup\mathcal{B}_{r+1\rightarrow r}^{\tau}\cup\left[\mathcal{F}_{r\rightarrow r+1}^{\tau-1} \setminus \left(\mathcal{F}_{r-1\rightarrow r}^{\tau-1}\cup\mathcal{B}_{r\rightarrow r-1}^{\tau-1}\right)\right];$\\update: $D(\tau+1)=D(\tau+1)+|\mathcal{D}_{r}^{\tau}|$;}
\ENDFOR

\STATE{\label{ln:betweenfors}detection: $\mathcal{D}_{1}^{\tau}=\mathrm{ISIC}(\mathcal{B}_{2\rightarrow 1}^{\tau} )$;\\update: $D(\tau+1)=D(\tau+1)+|\mathcal{D}_{1}^{\tau}|$;\\initialize: $\mathcal{F}_{1\rightarrow2}^{\tau}\equiv\mathcal{D}_{1}^{\tau}$;}

\STATE{\label{ln:back}\underline{Forward message-passing:}}
\FOR {$r = 2\to \bar{r}-1$}
\STATE{\label{ln:forwards}detection: $\mathcal{D}_{r}^{\tau}=\mathrm{ISIC}(\mathcal{F}_{r-1\rightarrow r}^{\tau})$;\\forward message:\\$\mathcal{F}_{r\rightarrow r+1}^{\tau}= \mathcal{D}_{r}^{\tau}\cup\mathcal{F}_{r-1\rightarrow r}^{\tau}\cup\left[\mathcal{B}_{r\rightarrow r-1}^{\tau} \setminus \left(\mathcal{B}_{r+1\rightarrow r}^{\tau}\cup\mathcal{F}_{r\rightarrow r+1}^{\tau-1}\right)\right];$\\update: $D(\tau+1)=D(\tau+1)+|\mathcal{D}_{r}^{\tau}|$;}
\ENDFOR

\STATE{\label{ln:iterwhile}detection: $\mathcal{D}_{\bar{r}}^{\tau+1}=\mathrm{ISIC}(\mathcal{F}_{\bar{r}-1\rightarrow \bar{r}}^{\tau})$;\\initialize: $\mathcal{B}_{\bar{r}\rightarrow\bar{r}-1}^{\tau+1}\equiv\mathcal{D}_{\bar{r}}^{\tau+1}$;\\update: $D(\tau+1)=D(\tau+1)+|\mathcal{D}_{\bar{r}}^{\tau+1}|$;\\update: $\tau=\tau+1$;}
\ENDWHILE
\\
\footnotesize{\vspace{2mm}Note: the set difference operation $\setminus$ in the generation of $\mathcal{B}_{r\rightarrow r-1}^{\tau}$ and $\mathcal{F}_{r\rightarrow r+1}^{\tau}$ avoids that the same message is unnecessarily repeated.}
\end{algorithmic}
\end{algorithm}

 \begin{figure}[tb]
\centering
\includegraphics[width= 0.8\textwidth]{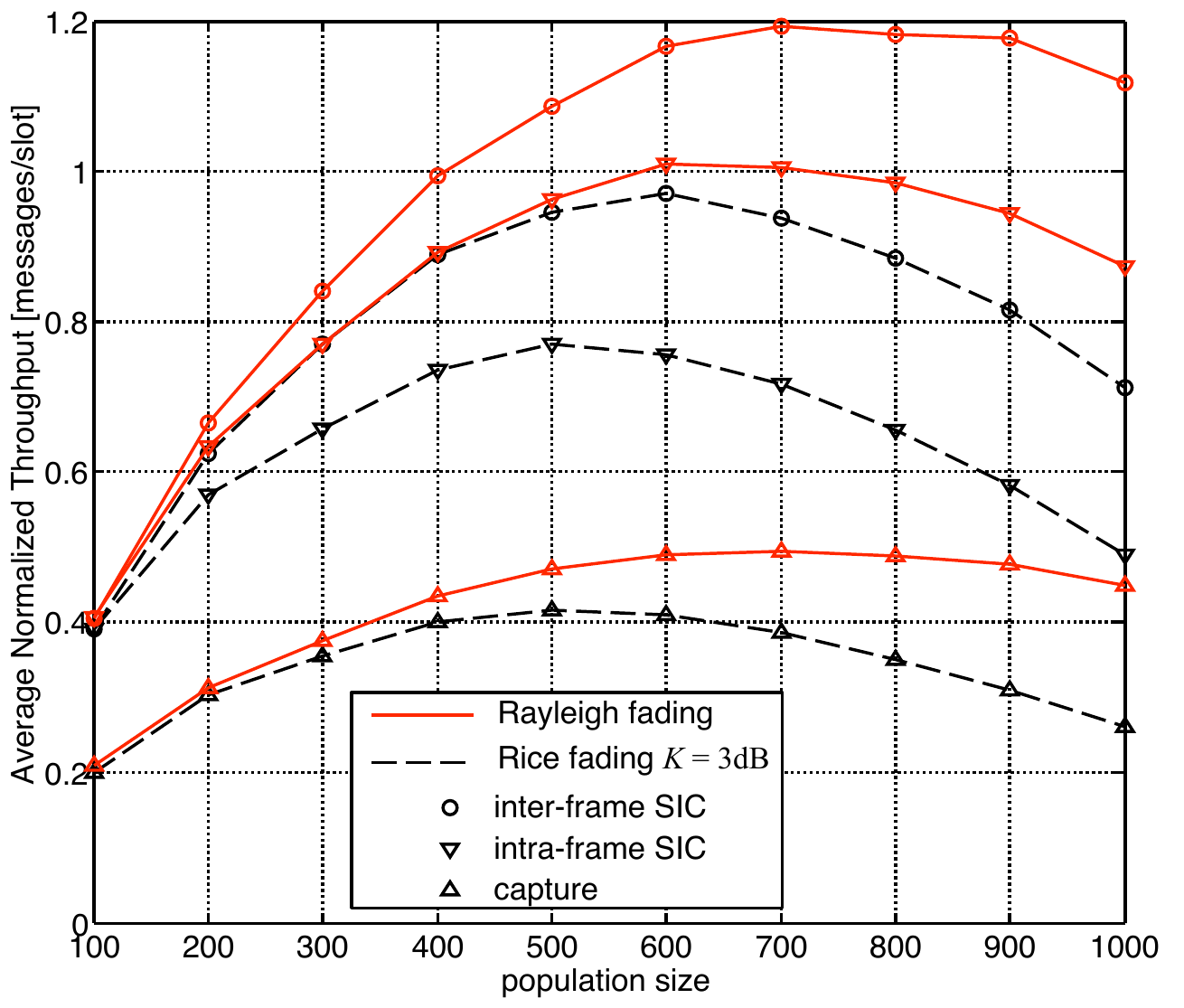}

\caption{Throughput vs. Tag Population Size.}
\label{fig:truput}
\end{figure}

 \section{Numerical Results}
We run a number of MATLAB$^\copyright$ simulations to compare quantitatively the performance of three PHY receiver structures in combination with  fixed-size Frame Slotted Aloha:
\begin{itemize}
\item[  I.] {\bf{Capture-only}} --- no Collision Recovery technique is adopted. However, as in real receivers,  the strongest received signal \emph{might} be succesfully decoded if the power gap with the remaining signals is sufficiently large.
\item[ II.] {\bf Intra-frame SIC} --- the classical SIC applied within each transmission slot.
\item[III.] {\bf Inter-frame SIC} ---  the exhaustive ISIC scheme based on Algorithm \ref{alg:ISIC}.
\end{itemize}
Note that only the latter option requires pseudo-random slot selection, while the previous two can be implemented with standard random selection. Our simulator reproduces the Gen2 modulation and coding format (ASK with Miller-8). An uniformly distributed random offset up to one symbol period is imposed to each Tag signal.
We run experiments with two channel models: Rayleigh fading and Rician fading with K-factor equal to 3dB. In both scenarios, the average channel power is the same for all tags with average Signal-to-Noise Ratio of 20 dB.  

We ran a set of simulations with fixed frame size $K=128$ and varying the tag population size $I \in [100,1000]$. For each experiment, we record the number $M$ of frames required to read all Tags. We define the average normalized throughput as the ratio $P \eqdef \frac{I}{K \cdot \mathbb{E}[{M}]}$. In Fig.\,\ref{fig:truput} we plot $P$ versus the population size $I$ for each receiver structure and channel type (total of six combinations).

As expected, the Rayleigh channel yields higher throughput than Rician for all three receivers.
This is due to the higher degree of  randomness in the received power,  which increases the probability of triggering radio capture\footnote{As a side note, we observed in the simulations that radio capture might occur also in case of signals colliding with equal power, provided that the time offset between the signals is large enough for the receiver to hook to the first one. In other words, time and power diversity both concur to trigger radio capture for signals with the given modulation format. {Furthermore,  we observed that the impact of channel estimation error on the SIC (and on the ISIC) is negligible for the considered operating conditions.}} and, in turn, successful  SIC.
The classical intra-frame SIC doubles the throughput compared to the simple capture-only receiver.
Compared to classical SIC, the proposed ISIC scheme brings a further throughput gain of 20\% - 25\% depending on the channel. 
Recall that such gain was obtained without any modification of the standard PHY layer format.

 \begin{figure}[tb]
\centering
\includegraphics[width= 0.8\textwidth]{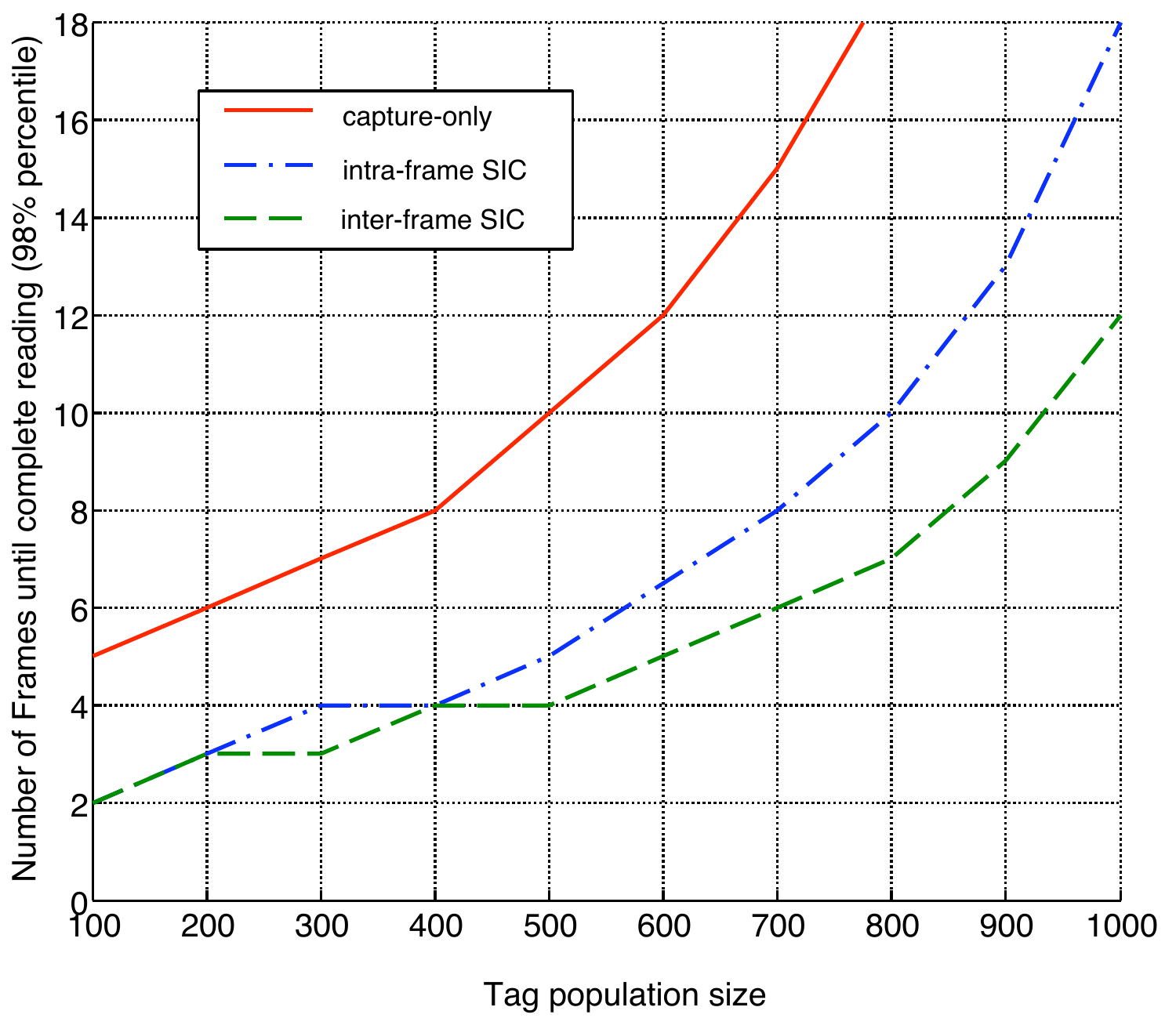}

\caption{Reading time: 98-percentile of the number of frames in a complete Reading Cycle vs. Tag population size for Rician channel. }
\label{fig:delay}
\end{figure}

A by-product of the ISIC gain is that the throughput remains high even if the size of the Tag population exceeds considerably the theoretical optimal value for the given frame size --- that is however unknown for arbitrary receiver structure. In other words, with ISIC the overall system is less sensitive to the size of Tag population, thus reducing the need for an accurate initial estimation.

For a given Tag population and frame size, an increase in throughput  translates into shorter reading cycles. In Fig.\,\ref{fig:delay} we plot the 98th percentile of the Reading Cycle size, i.e. the number $M$ of frames required to read all Tags, for the Rician channel. The plot shows that ISIC yields reading cycles shorter by up one third compared to the classical intra-frame SIC, with a dramatic improvement over the simple capture-only receiver. 

 \begin{figure}[tb]
\centering
\includegraphics[width= 0.8\textwidth]{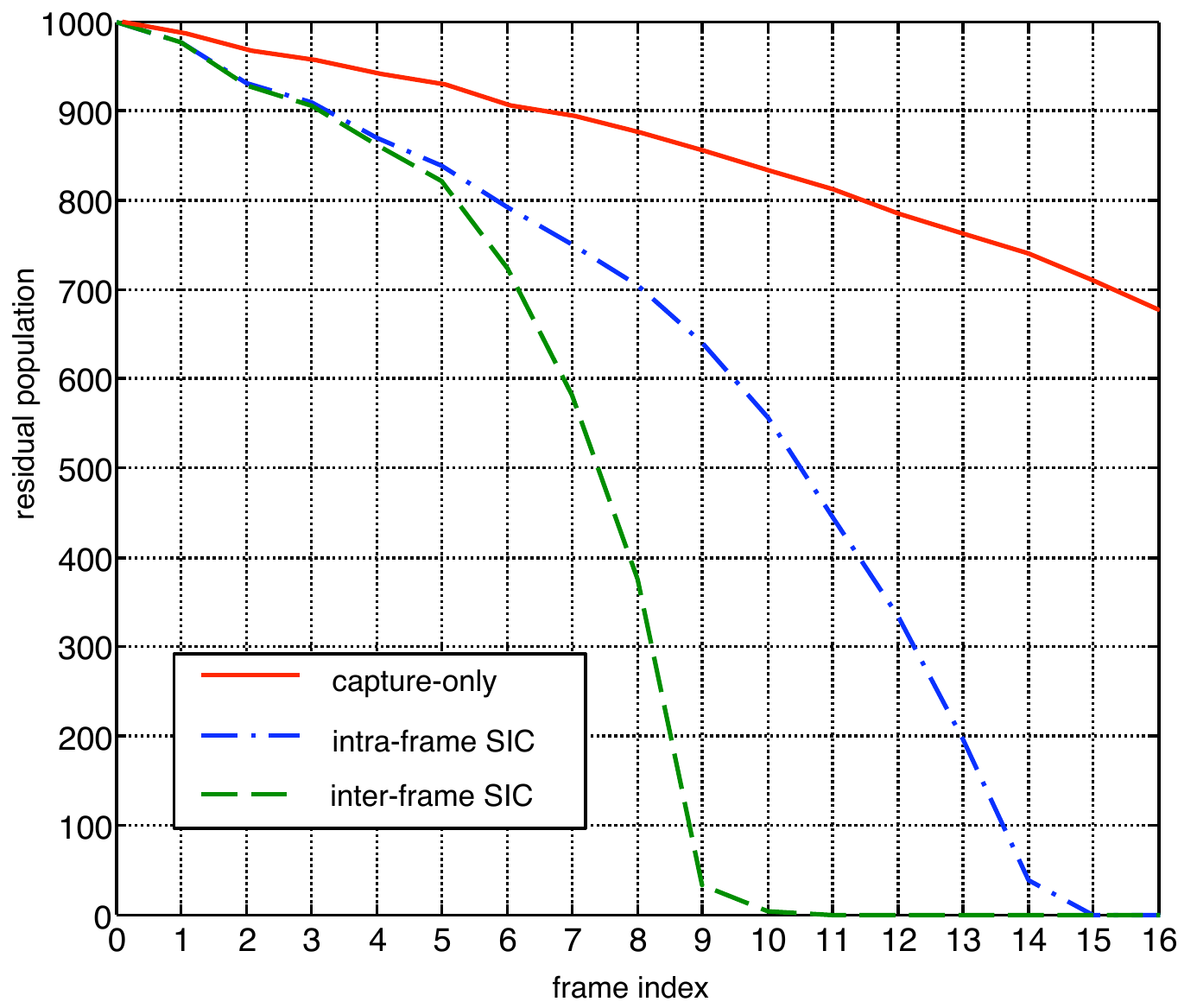}

\caption{Number of residual (not yet decoded) Tags vs. frame index for $I=1000$ and Rician channel.  }
\label{fig:realizzazione}
\end{figure}

To illustrate further, in Fig.\,\ref{fig:realizzazione} we show the number of residual Tag messages (not yet decoded) versus the frame index $r$ for a sample realization  with $I=1000$ and Rician channel.
Comparing the classic SIC and ISIC curves, it can be seen that inter-frame cancellation starts to become effective after a few iterations --- the knee point is at $r=5$ in this specific realization--- and from that point onwards it causes the  incremental reading rate to double with respect to the classical SIC (compare the two slopes). This suggests that in those RFID applications where the Tag population is dynamically churning (continuous arrivals), ISIC gain can be higher than in the static case.

\section{Conclusions and Future Work}

In this preliminary work we have shown that a simple pseudo-random (as opposite to random) access method can immediately enable inter-frame signal cancellation in a RFID system \emph{with standard Gen2 compliant PHY modulation format}.
We believe that the proposed protocol is just the most obvious representative of a wider class of  access protocols that are still to be explored by the research community, where pseudo-random access strategy are used in combination with advanced  PHY receiver structures.

We have so-far exploited pseudo-randomization to identify \emph{past} slot positions for the purpose of implementing \emph{backward} inter-frame signal cancellation. In the future we will study the potential leveraging of pseudo-randomization in the \emph{forward} direction, and specifically to support the inference of \emph{future} slot position as needed to implement some form of Soft Combining. Our preliminary investigations along this direction show however that this approach is not rewarding as far as the legacy Gen2 modulation  format is retained. We have instead observed that significant gains of pseudo-random access in the forward direction can be achieved by deploying more powerful error correcting codes, but that implies a major change of the  PHY layer Gen2 standard.

In the progress of our work we are planning to prototype the pseudo-random Aloha scheme in Gnu Radio \cite{donno} in order to carry out real-world testing.


\bibliographystyle{ieeetr}
\bibliography{rfidbiblio}

\end{document}